\documentclass[12pt]{iopart}
\usepackage[utf8]{inputenc}
\usepackage{graphicx}
\usepackage{hyperref}
\usepackage[usenames,dvipsnames]{xcolor}		
\usepackage{siunitx}
\expandafter\let\csname equation*\endcsname\relax
\expandafter\let\csname endequation*\endcsname\relax
\usepackage{amsmath}
\usepackage{textcomp}
\usepackage{booktabs}
\usepackage[version=4]{mhchem}

\hypersetup{
   pdffitwindow=true,
   colorlinks=true,  
   breaklinks=true,  
   urlcolor= NavyBlue,   
   linkcolor=WildStrawberry,    
   citecolor=blue,   
   filecolor=magenta,
   menucolor=black,  
   pdftitle={Comment on Collision and radiative processes in emission of atmospheric carbon dioxide},     
   pdfauthor={Mario Lino da Silva},
}

\PassOptionsToPackage{hyphens}{url}\usepackage{hyperref}
\usepackage{url}
\makeatletter
\g@addto@macro{\UrlBreaks}{\UrlOrds}
\makeatother
\usepackage{setspace}

\begin{document}

\title[]{Comment on ``Collision and radiative processes in emission of atmospheric carbon dioxide''}

\author{M. Lino da Silva and J. Vargas}

\address{Instituto de Plasmas e Fus\~{a}o Nuclear, Instituto Superior T\'{e}cnico, Universidade de Lisboa, Av. Rovisco Pais 1, 1049-001, Lisboa, Portugal}
\ead{mlinodasilva@tecnico.ulisboa.pt}
\vspace{10pt}
\begin{indented}
\item[]December 2019
\end{indented}

\begin{abstract}
Recently, Smirnov published a paper (B. M. Smirnov, "Collision and radiative processes in emission of atmospheric carbon dioxide", 2018, \emph{J. Phys. D.: Appl. Phys.}, Vol. 51, No. 21, pp. 214004) which dismisses the role of increasing concentrations of anthropogenic \ce{CO2} on global warming of planet Earth. We show that these conclusions are the consequence of two flaws in Smirnov's theoretical model which neglect the effects of the increased concentrations of \ce{CO2} on the absorption of Earth's blackbody radiation in the 12--15$\mu$m region. The influence of doubling the concentration of \ce{CO2} in the atmosphere on the surface temperature is not $\Delta T=0.02\si{\kelvin}$, or even $\Delta T=0.4\si{\kelvin}$ if only one of the two mistakes in Smirnov's analysis is corrected. The correct value lies within $\Delta T=1.1-1.3 \si{\kelvin}$ as outlined by other authors analysis using simplified, yet more theoretically consistent models.
\end{abstract}

%
%
%
%
%

\section*{Foreword}
\small
\footnotesize
\begin{singlespace}
Originally, this comment to the article: B. M. Smirnov, "Collision and radiative processes in emission of atmospheric carbon dioxide", 2018, \emph{J. Phys. D.: Appl. Phys.}, Vol. 51, No. 21, pp. 214004, was submitted to \JPD No agreement between the authors and the editorial board could be achieved regarding the proper size and structure to the comment to this article, and as such the authors decided not to pursue publication in \emph{J. Phys. D.: Appl. Phys.}, instead publishing the comment in its original form in arXiv. A few suggestions from the editorial board from \JPD were nonetheless integrated in this final version of our comment to Smirnov's article.
\end{singlespace}
\normalsize

\section*{Introduction}

The authors have been pointed out to this article by Smirnov which claims anthropogenic \ce{CO2} sources to be negligible as regarding global warming effects. This is a surprising conclusion that directly contradicts a large amount of research in the last decades, regarding the topic of global warming. Upon close inspection we have been able to identify two different mistakes in the methodology of the work that lead to this flawed conclusion.

Upon this first review, and after our decision to write a rebuttal of the article conclusions, we felt that a more extensive comment was warranted: Providing a summary of the major works regarding the monitoring of energy exchange processes at global Earth level would add a better insight on the consensus that has been formed at the scientific community level regarding the impact of anthropogenic \ce{CO2} sources on climate change. Since this is a topic with very high impact societal impact, we hope to provide a summary of the major findings regarding this topic, thereby unambiguously dispelling any misconceptions that may arise from works such as the one in review.

This comment to the article by Smirnov is split into three sections. Section \ref{sec:summary}
provides a summary on the energy budget of the Earth, as determined by detailed measurements over the last decade. Section \ref{sec:Smirnov} examines in detail the shortcomings of Smirnov's work, and Section \ref{sec:application} presents a model that relies on the same initial data as Smirnov, but yields the correct conclusions, as it follows a more appropriate methodology. The reader with less available time will still be able to quickly refer to Section \ref{sec:Smirnov} to understand the flaws of Smirnov's work, which is the main objective of this comment.

\section{Earth's Energy Budget: A Short Summary}
\label{sec:summary}

%
%

\begin{figure*}[htbp]
\centering
\includegraphics[width=0.8\textwidth]{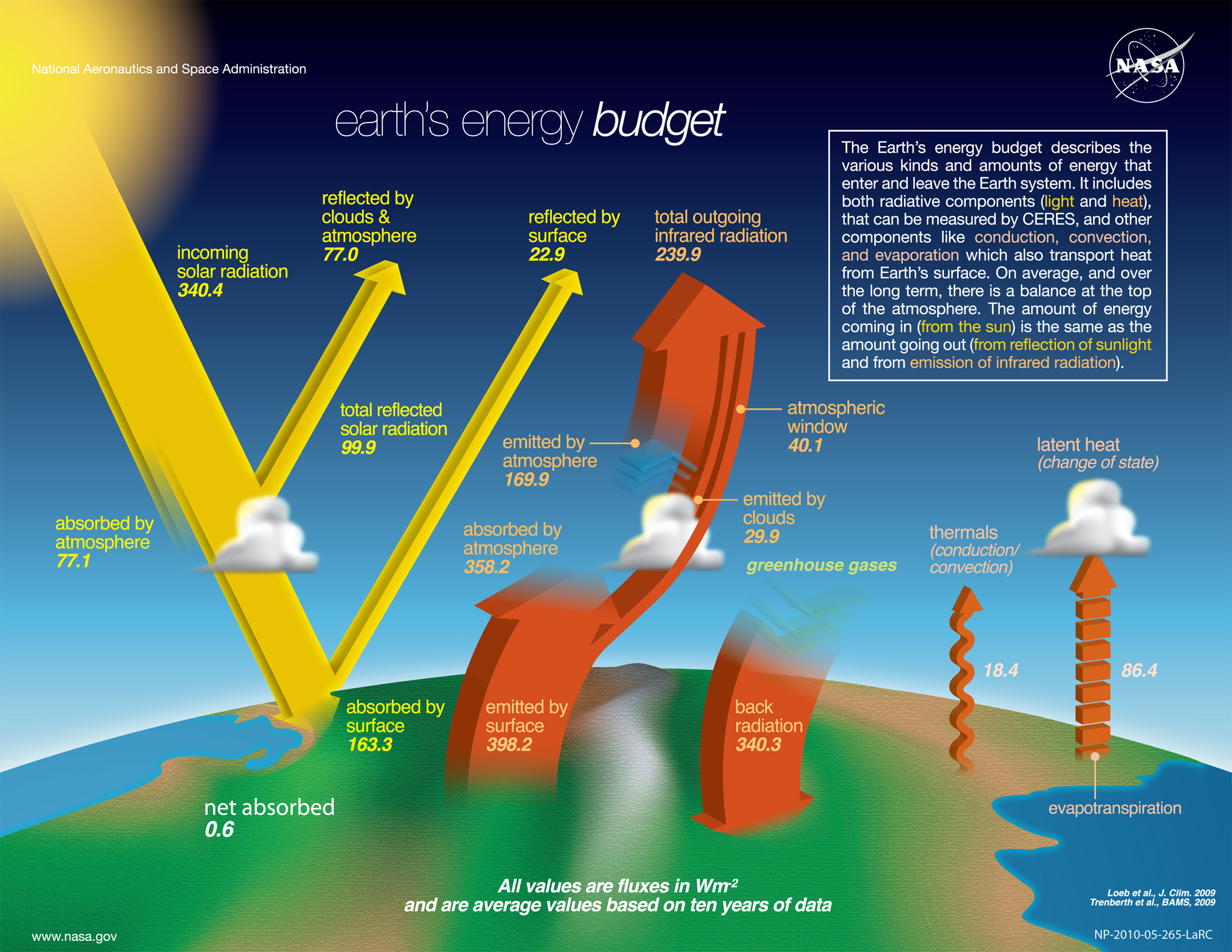}
\caption{Earth's energy budget \cite{Earthradfluxes}}
\label{fig:earthradfluxes}
\end{figure*}

The values of the heat fluxes for these different energy exchange processes are known to good accuracy, as these have been determined from the average of ten years of recorded data \cite{Loeb:2009,Tremberth:2009}. These are presented in Fig. \ref{fig:earthradfluxes}

If this equilibrium is disrupted (like for example when the opacity of the atmosphere increases due to anthropogenic factors or other), the energy flux imbalance (in this example due to less energy escaping to Space when compared to the energy from the Sun irradiation) will lead to a new equilibrium state, with different temperatures. This change is not instantaneous, owing to the large thermal inertia from the Earth (mostly due to the large volume of its oceans), and may take many years before a new stable thermodynamic equilibrium is reached. In the 2005--2010 year interval, this imbalance was estimated to be around 0.6\si{\watt\per\metre\squared} \cite{Hansen:2011}.


\subsection{Atmospheric opacity and its influence on Earth's energy balance}

Here we shortly discuss the downward solar radiative fluxes and the associated losses in the path from the edge of the atmosphere towards the surface, and the outward Earth surface and atmospheric radiative fluxes towards Space.

\begin{figure}[htbp]
\centering
\includegraphics[width=0.7\textwidth]{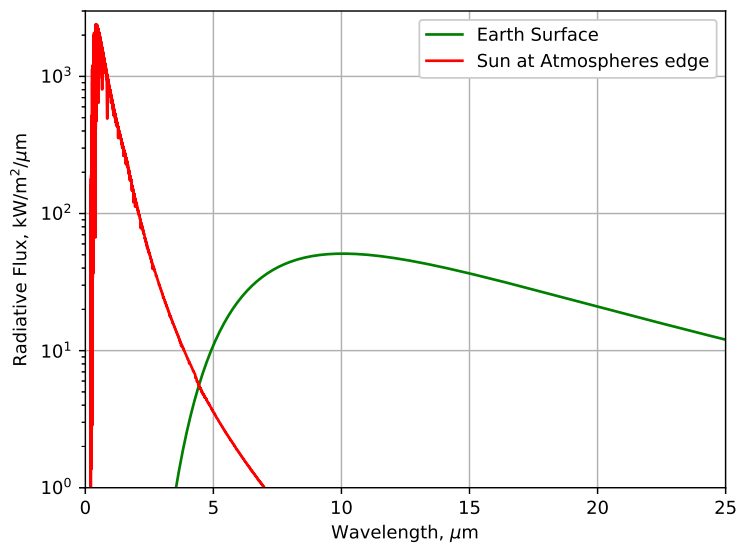}
\caption{Comparison between the downward Solar radiative fluxes and the upward Earth radiative fluxes}
\label{fig:solearthfluxes}
\end{figure}

Solar fluxes follow closely a Planck blackbody distribution with a characteristic temperature around 5,780\si{\kelvin}. 98\% of the radiative power is emitted in the 0.25--4 \si{\micro\metre} range. Radiative emission from the Earth's surface follows a Planck blackbody distribution with a characteristic temperature of 288\si{\kelvin}. 98\% of the radiative power is emitted in the 5--80 \si{\micro\metre} range. Finally, atmospheric emission fluxes do not follow a blackbody distribution, but are limited by a Planck blackbody at a characteristic temperature of 217\si{\kelvin} in the tropopause and 287\si{\kelvin} in the boundary layer region \cite{Bianchini:2003}. One first important remark is that both inbound Solar fluxes and outbound Earth fluxes have essentially no overlap spectrally-speaking. A comparison of both radiative fluxes is presented in Fig. \ref{fig:solearthfluxes}.

\begin{figure}[htbp]
\centering
\includegraphics[width=0.7\textwidth]{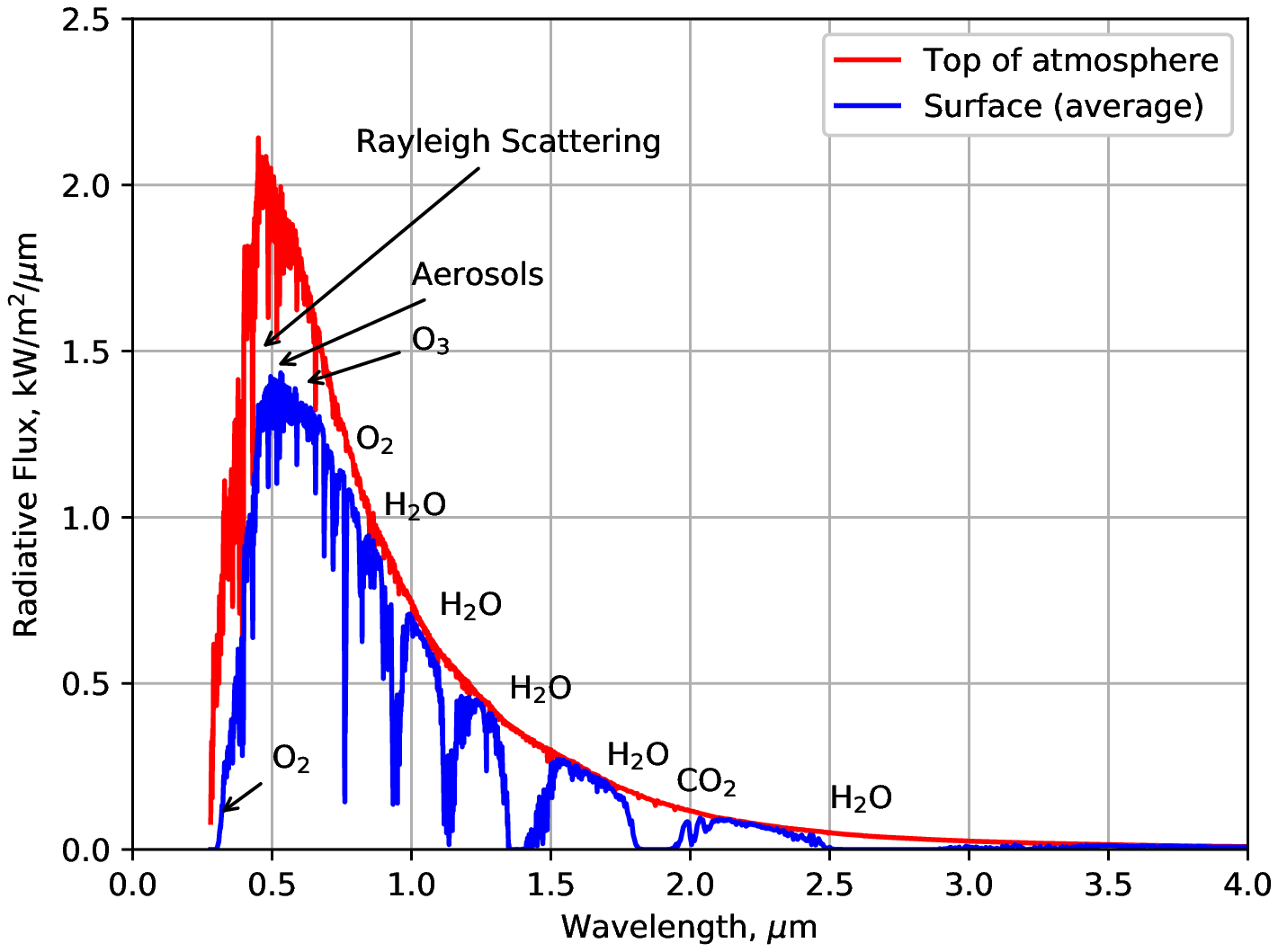}
\includegraphics[width=0.7\textwidth]{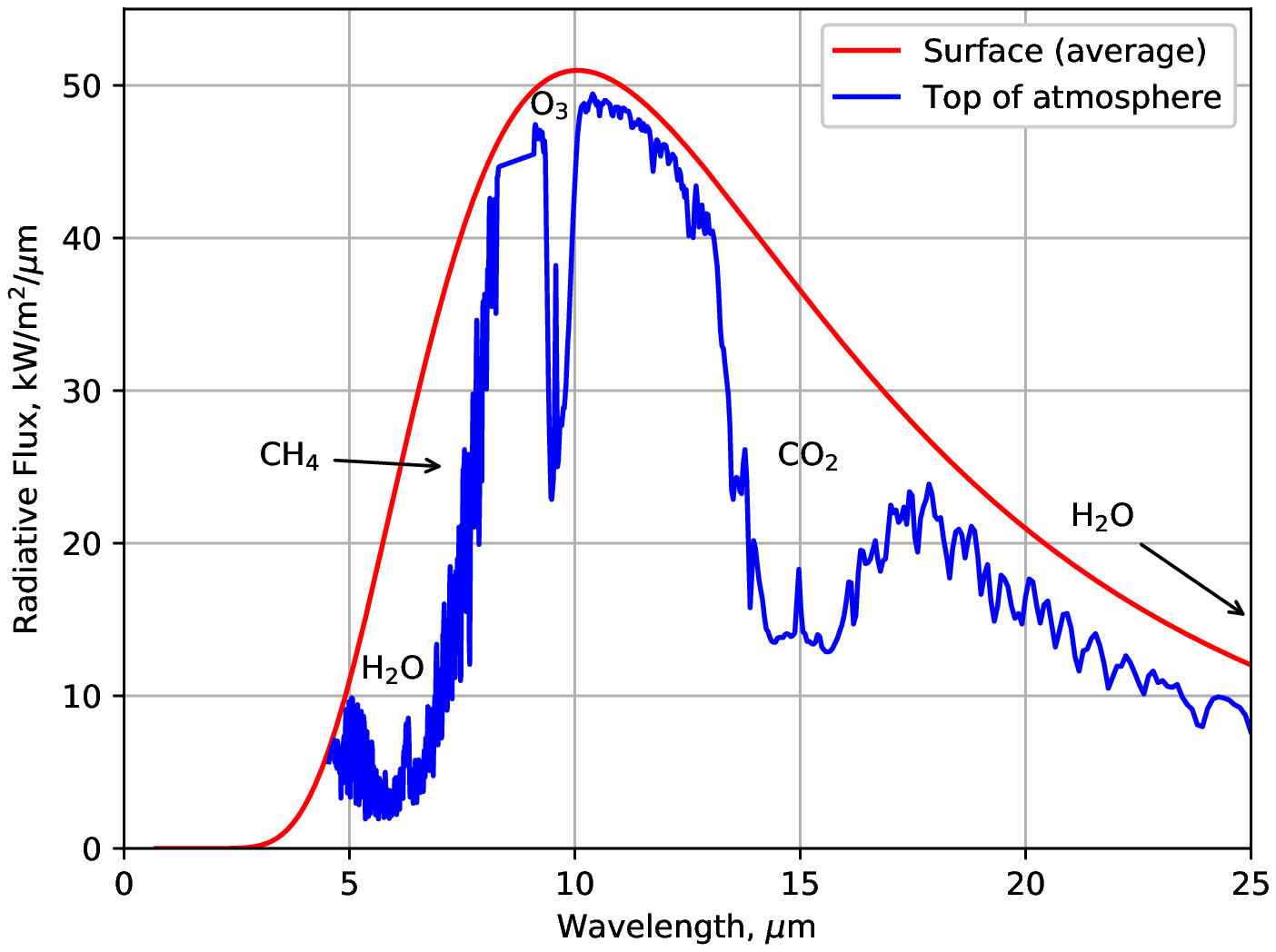}
\caption{Downward Solar radiative fluxes at atmosphere edge and surface (top figure), and upward Earth radiative fluxes at surface and atmosphere edge (bottom figure)}
\label{fig:fluxes}
\end{figure}

Atmospheric opacity, alongside with scattering processes, only allows a fraction of the Solar radiation to reach Earth's surface, and only a fraction of the Earth's radiation to escape to Space. To illustrate this we have considered the 1976 Standard Atmosphere, and the data from Ref. \cite{SolarSpectrum} to show the spectral distribution of the downward Solar fluxes at the edge of the atmosphere and Earth's surface. For the upward infrared blackbody radiation, a radiative transfer simulation with the MODTRAN online tool has been carried out to obtain the upward spectra\footnote{Note that this corresponds for the contribution from the blackbody Earth fluxes that are not absorbed by the atmosphere, and the contributions from the atmosphere itself, modeled with strata of different temperatures} at the edge of the atmosphere \cite{MODTRAN}. In the simulation parameters, contemporary concentrations for \ce{CO2}, \ce{CH4} and \ce{O3} were considered. The corresponding radiative fluxes are presented in Figs. \ref{fig:fluxes}, on the top the downward solar radiative flux and on the bottom the upward Earth radiative flux.

As it is well known, a significant amount of the VUV-Visible radiation from the Sun is absorbed by the Ozone layer, or scattered by different processes. \ce{O2}, a major constituent of the air, also absorbs in the VUV through photodissociation processes (Schumann--Runge and Schumann--Runge continuum), and through a few rovibronic transitions in the visible region. Then in the near-IR and IR region, water vapour is the main absorber of radiation, with a near-negligible contribution by the 2.7\si{\micro\meter} band of \ce{CO2}. For the Earth surface and atmosphere infrared radiation, the picture is significantly different. Homonuclear molecules such as \ce{N2} and \ce{O2} have no permanent dipole moment, and as such may not have any rovibrational transitions, and do not absorb at all in the IR-MWIR region. \ce{H2O} is again a large absorber of radiation, however \ce{CO2} is now responsible for a large gap in the transmissivity of such radiation towards Space, centered around 15\si{\micro\meter}. Then there are some minor contributions from \ce{O3} and \ce{CH4}.

\section{Review of the Smirnov Analysis}
\label{sec:Smirnov}

Following the outline for the different heat fluxes for the Earth energy budget, as determined from extensive measurements by other authors, it is now adequate to proceed to a detailed review of the analysis presented by Smirnov. We may show that it is essentially flawed in two key points. These are discussed in detail in this section:

The first flaw lies in the atmospheric energy balance that is proposed. The key equation that summarizes this balance in Smirnov's paper (eq. 6.17) is reproduced below

\begin{align}
\frac{1}{S}=&\frac{4J_E}{T_E}-\frac{4\left[J_\downarrow-J\left(\textrm{CO}_2\right)\right]}{T_\downarrow}\left(1-\frac{h_\downarrow}{h_0}\right)\notag\\
&-\frac{J\left(\textrm{CO}_2\right)}{T\left(\textrm{CO}_2\right)}\left[1-\frac{h\left(\textrm{CO}_2\right)}{h_0}\right]+\frac{J_c}{\delta T}
\label{eq:Smirnov}
\end{align}

where $S$ is the climates sensibility parameter, $h_0$ is the altitude of the troposphere, $h_\downarrow$ the altitude of the column where the atmosphere radiates significantly, $h(\textrm{CO}_2)$ the altitude of the column where \ce{CO2} radiates significantly, $T_E$ is the Earth's temperature, $T_\downarrow$ the radiative temperature of the atmosphere, $T(\textrm{CO}_2)$ is the radiative temperature of \ce{CO2}, $\delta T$ is the temperature difference between the Earth's surface and tropopause temperatures, $J_E$ is the Earth's blackbody heat flux, $J_c$ is the convective heat flux from Earth's surface to the atmosphere,$J_\downarrow$ is the overall radiative heat flux from the atmosphere to Earth's surface, $J(\textrm{CO}_2)$ is the radiative heat flux from atmospheric \ce{CO2} to Earth's surface.

Here the problem lies in Earth's blackbody energy flux $J_E$. Although the value for this flux (386\si{\watt\per\metre\squared}) lies close to the accepted value (398\si{\watt\per\metre\squared}; see Fig. \ref{fig:earthradfluxes}), there is no accounting for the absorption of this radiation from the atmosphere (mostly from \ce{H2O}, \ce{CO2}, and \ce{CH4}). In reality, most of this blackbody radiation is absorbed by the atmosphere, with some of the re-emitted radiation escaping to Space (one may again refer to Fig. \ref{fig:earthradfluxes} for the corresponding fluxes). As such, treating $J_E$ as a constant term will significantly underpredict the sensitivity of the heat fluxes balance of Eq. \ref{eq:Smirnov}, since the atmospheric concentrations of \ce{CO2} will significantly influence how much of this radiation will be absorbed by Earth's atmosphere. Other energy balance analysis have been carried out by Wilson \cite{Wilson:2012}, properly accounting for this effect, which yield values of $\Delta T=1.1-1.3$, much higher than the value proposed by Smirnov ($\Delta T=0.4\pm0.1$). A simple calculation will be presented in section \ref{sec:application} to illustrate the effects of \ce{CO2} on the absorption of this radiative flux.

The second major flaw of Smirnov analysis lies with the quantitative accounting for the contribution of anthropogenic \ce{CO2} (as opposed to natural \ce{CO2} sources) in the temperature increase analysis.

Smirnov does not cite a source for its estimation of 5\% contribution of anthropogenic sources for the total fluxes of \ce{CO2}. Nevertheless, this value is more or less equivalent to the ratio for the emission of anthropogenic \ce{CO2} to the one by natural sources (respiration of plants and animals and plant consumption by animals, plus out-gassing by oceans). Using the aforementioned values from Ref. \cite{Ciais:2014}, we obtain $32.6/(435+288)=4.5\%$.

This said, one may not simply claim that the contribution from the combustion of fossil fuels
to the Equilibrium Climate Sensitivity (ECS) is 5\% since this simply neglects the \ce{CO2} sink values from photosynthesis (-450Gt/yr) and from the Ocean (-294Gt/yr). The key effect of anthropogenic pumping of \ce{CO2} into the atmosphere is to disrupt the balance between source and sink terms of \ce{CO2}, and the aforementioned percentage is irrelevant since in a first approach (excluding any feedback mechanisms), doubling/halving the natural source/sink terms would significantly change the assumption by Smirnov (eq. 6.20) without actually changing the net growth of \ce{CO2} in the atmosphere (4Gt/yr), and hence the actual temperature increases.

Instead, any doubling of \ce{CO2} in the atmosphere has by all evidence to be assumed as the effect of anthropogenic sources since firstly, concentrations of \ce{CO2} have increased from about 300ppm in 1950, when fossil fuel burning became the major anthropogenic source of atmospheric \ce{CO2}\footnote{Compared to land use changes}, up to 410ppm today, a higher than 1/3 increase \cite{Ciais:2014,Hellevang:2015}, and secondly, only about 57\% of human-emitted \ce{CO2} is removed by the biosphere and oceans \cite{Canadell:2007,LeQuere:2009,Huang:2012}. Therefore the correct value to use for the temperature increase would rather be the one from eq. 6.19 if this wasn't also wrong (although much closer to the right value).

The authors again forward the interested reader towards the article from Wilson \cite{Wilson:2012} which elegantly outlines several approaches with different levels of complexity for solving this problem, also outlining several well known feedback effects in climate change modeling.


\section{A Quick Illustration of the Contribution of CO\texorpdfstring{$_2$ }\ to Atmosphere Radiative Forcing}
\label{sec:application}

As discussed before, the amount of Earth's blackbody radiation that is absorbed by the atmosphere is very sensible to variations in \ce{CO2} atmospheric concentrations. To demonstrate this, we provide a simple testcase, using the same environmental parameters than Smirnov (essentially we use the average surface  $T_E$ and air $T_\downarrow$ temperatures, and column height $h_\downarrow$). For $J_E$ (flux from the Earth) we consider a Planck blackbody at 288\si{\kelvin}, same as Smirnov.

\begin{figure}[htbp]
\centering
 \includegraphics[width=0.7\textwidth]{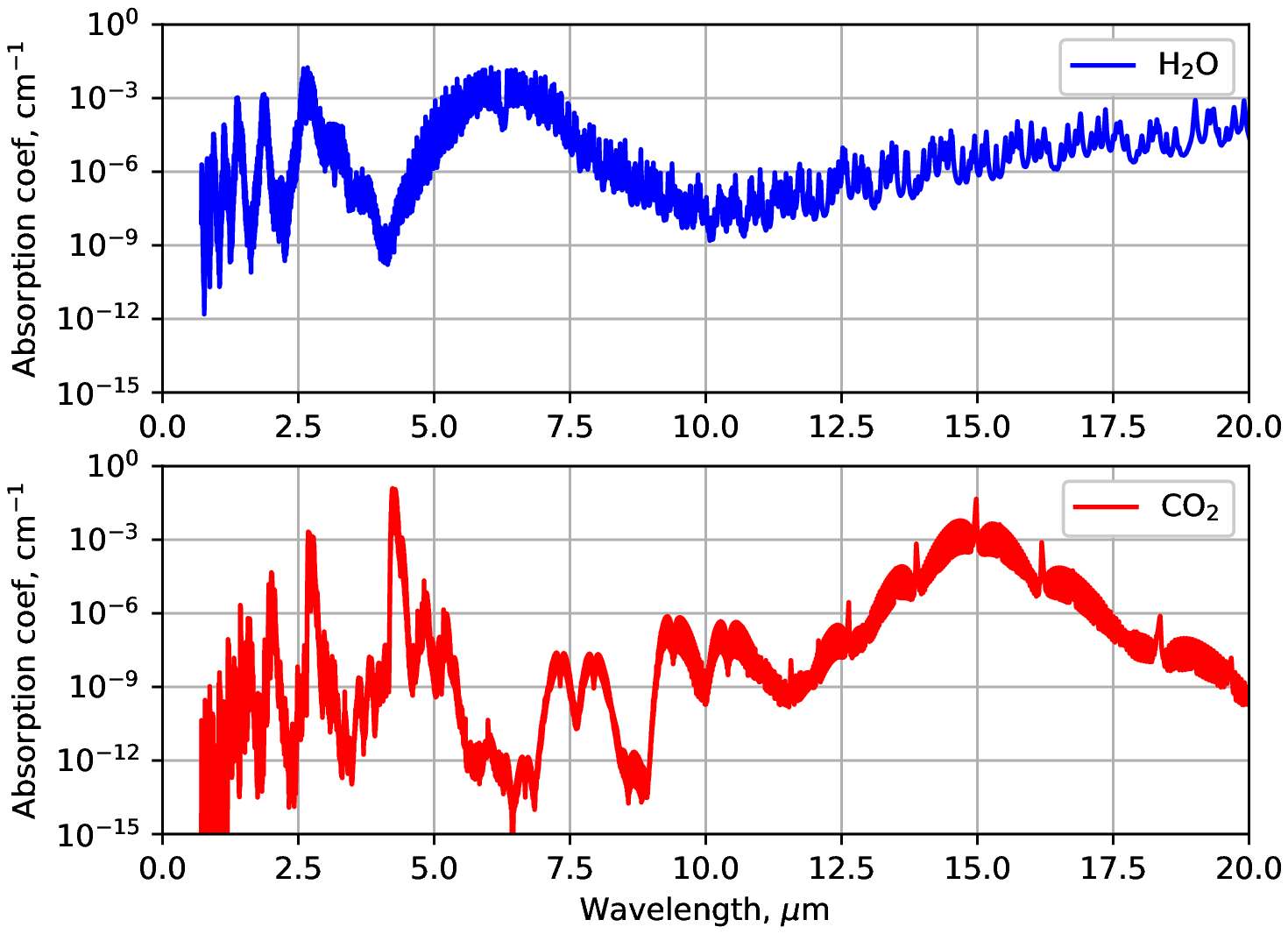}
  \caption{Absorption coefficient for 0.25\% \ce{H2O} and 409ppm \ce{CO2} at atmospheric pressure}
 \label{fig:abscoCO2H2O}
\end{figure}

Absorption coefficients for \ce{CO2} (and \ce{H2O}\footnote{For the sake of simplicity we neglect other minor contributor to the greenhouse effect such as \ce{CH4} and \ce{O3}}) at $T_\downarrow=274\si{\kelvin}$ are retrieved from HITRAN on the Web (HotW) \cite{HITRANonline}. HotW provides very accurate air-broadened Lorentz half-widths for each absorption line, which is of the upmost importance when calculating absorption of the radiation from Earth, as the absorption from the wings controls the size of the spectral transmission window for this radiation. To obtain the exact absorption coefficients, we then multiply the obtained absorption coefficients for the molar fractions of \ce{CO2} and \ce{H2O} into consideration. We select the contemporary concentrations of \ce{CO2} (410ppm) and an averaged value of 0.25\% for the water vapour concentration in the air, as proposed in Ref. \cite{Lodders:2015}. This way we get the correct line peaks, and the correct broadening (Lorentz at 1 atmosphere). The absorption coefficients are calculated with a continuous grid, for the smallest step proposed by the online database for both spectra (0.01\si{\centi\metre^{-1}}), so as to achieve good accuracy in the modeling of the line profiles. To further enforce this, we set a line cutoff calculation criteria of 100 line full widths at half maximum (FWHM). The obtained spectra are presented in Fig. \ref{fig:abscoCO2H2O}.

Radiative transfer calculations are then carried over the column height using Beer-Lambert's Law:

\begin{equation}
I_{Space}=B_{\nu,T=288K}\exp(-\alpha' h_\downarrow)
\end{equation}

Where $I_{Space}$ is the radiation that escapes to Space and $\alpha'$ is the absorption coefficient corrected for self-absorption $\alpha'=\alpha\left[1-\exp(-1.4388\nu/T_\downarrow)\right]$ and $\nu$ is the wavenumber in \si{\centi\metre^{-1}}.

Here we have considered five cases regarding the absorption coefficient, with the total absorption coefficient being respectively $\alpha=[\alpha(\textrm{H}_2\textrm{O});\ \alpha(\textrm{CO}_2);\ 2\times \alpha(\textrm{CO}_2);\ \alpha(\textrm{H}_2\textrm{O})+\alpha(\textrm{CO}_2);\ \alpha(\textrm{H}_2\textrm{O})+2\times \alpha(\textrm{CO}_2)]$.

\begin{figure}[htbp]
\centering
 \includegraphics[width=0.7\textwidth]{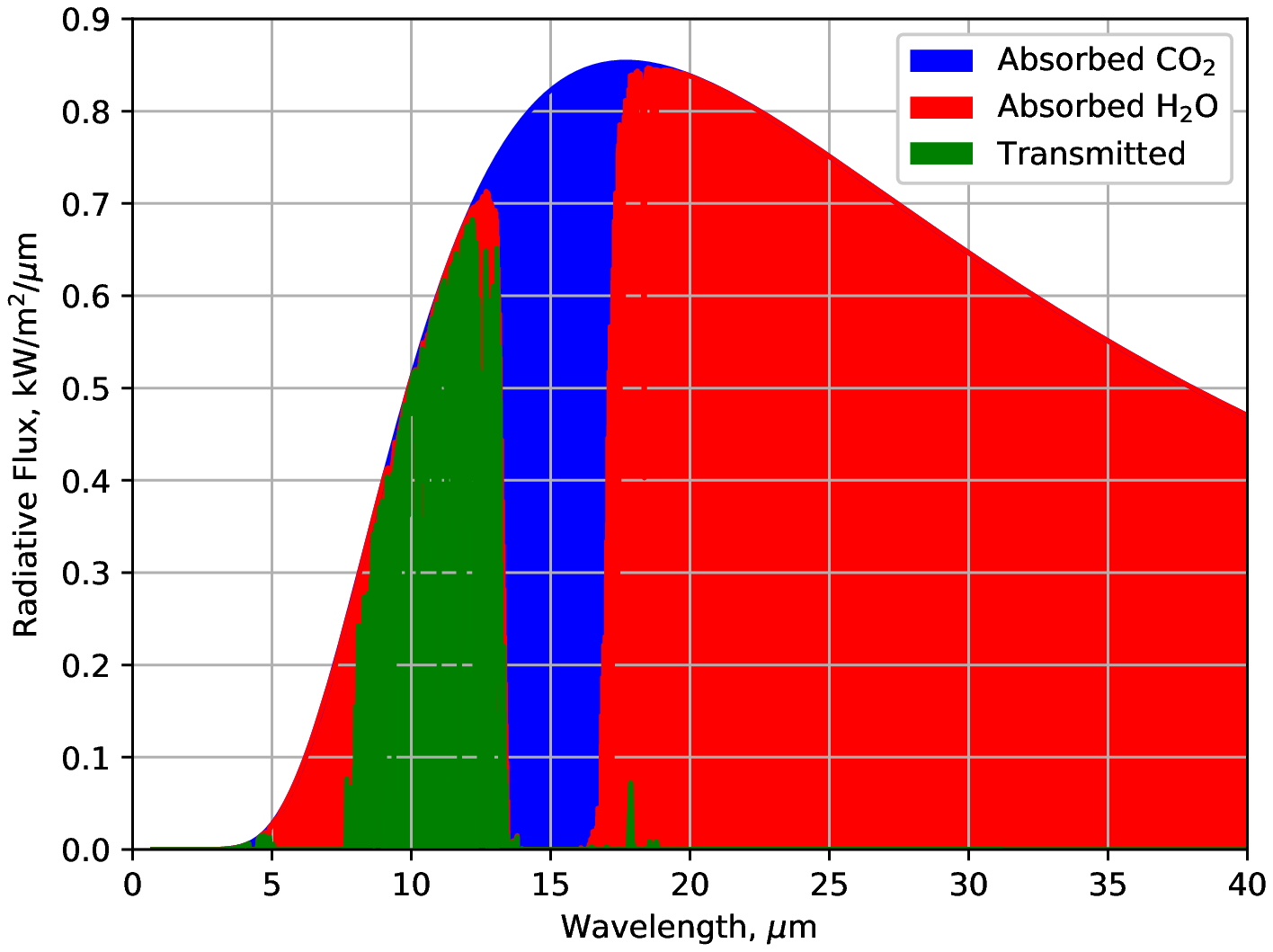}
 \includegraphics[width=0.7\textwidth]{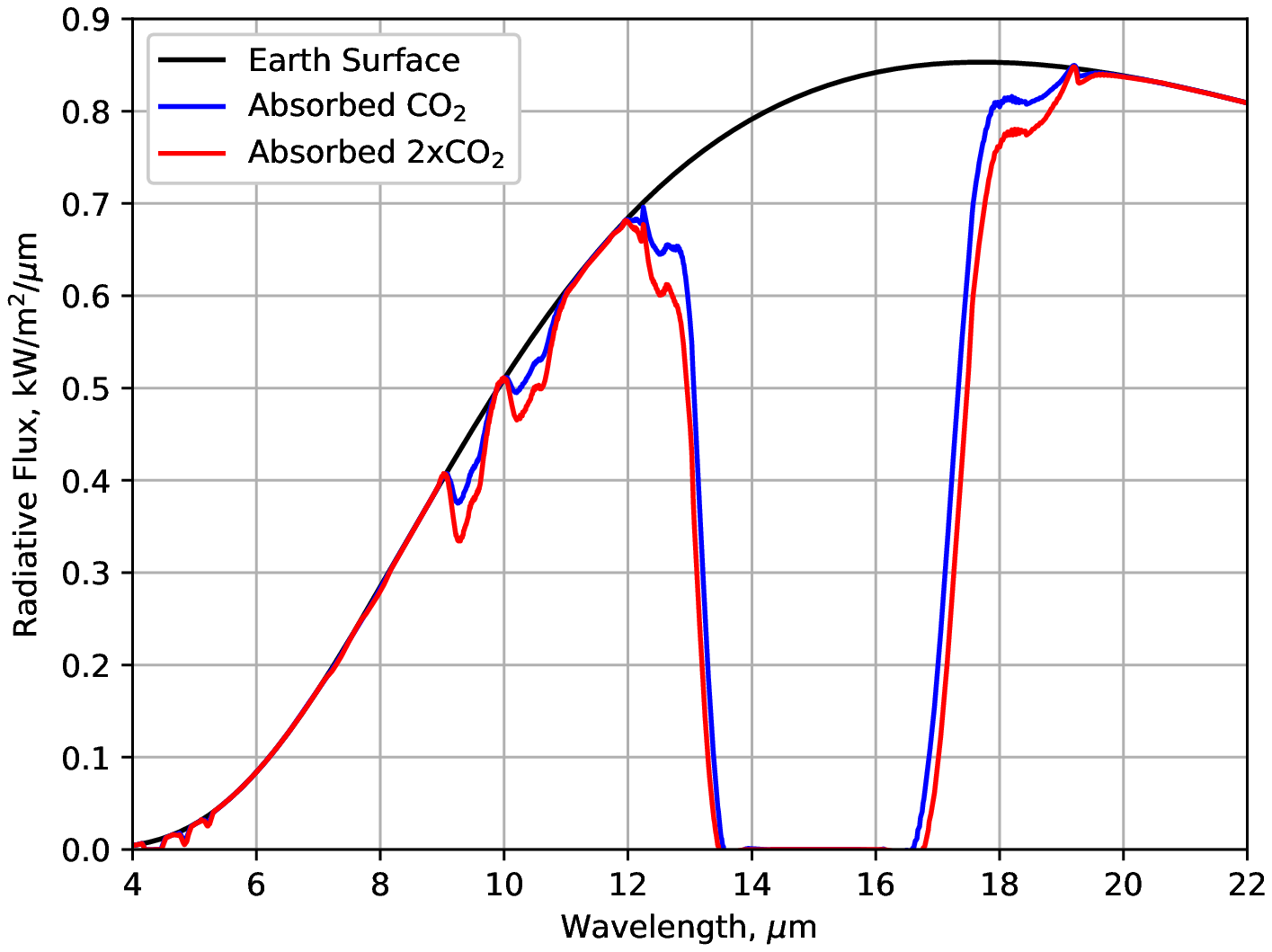}
  \caption{Transmitted Earth radiation towards Space (green); Absorbed by \ce{H2O} (blue) and absorbed by \ce{CO2} (red); top figure. Transmitted radiation towards Space for 409ppm \ce{CO2} and doubled concentration of \ce{CO2}; bottom figure}
 \label{fig:blackbody_transmitted}
\end{figure}

We first present the transmitted and absorbed radiation obtained for the contemporary concentrations of \ce{CO2} and \ce{H2O} (409ppm and 0.25\%) (Fig. \ref{fig:blackbody_transmitted}; top) and the differences in transmitted radiation, considering only absorption from \ce{CO2} at contemporary concentrations (409ppm), and the double for this value (818ppm) (Fig. \ref{fig:blackbody_transmitted}; bottom). Table \ref{tab:absorption} summarizes the losses simulated in this exercise (again, excluding any radiative emission from the atmosphere).

\begin{table}[!htbp]
\centering
\caption{Loss percentages (excluding contributions from the atmosphere)}
\label{tab:absorption}
\begin{tabular}{l c c c c c}
\toprule
\midrule
& $\textrm{CO}_2$ %
& $2\times\textrm{CO}_2$ %
& $\textrm{H}_2\textrm{O}$ %
& $\textrm{H}_2\textrm{O}+\textrm{CO}_2$ %
& $\textrm{H}_2\textrm{O}+2\times\textrm{CO}_2$\\%
\midrule
losses (\%) & 20.5\% & 22.5\% & 68.7\% & 74.4\% & 75.4\%\\
losses (\si{\watt/\metre\squared}) & 80.0 & 87.7 & 268.2 & 290.4 & 294.2\\
\midrule
\bottomrule
\end{tabular}
\end{table}

Here we can confirm the usual evidences of contemporary climate change models: Firstly \ce{H2O} is the main greenhouse gas responsible for blocking Earth infrared radiation from escaping to Space. However, \ce{CO2} is also seen to contribute with additional blocking of radiation around the 15\si{\micro\metre} spectral range. \ce{H2O} alone block's 68.7\% of Earth's radiation, whereas \ce{CO2} blocks 20.5\%. The two molecules block together around 74.4\% of Earth's radiation. Secondly, if we double the concentration of \ce{CO2}, we increase radiation blockage from 20.5\% to 22.5\% (2\% increase) if \ce{CO2} alone is considered, or from 74.4\% to 75.4\%  (1\% increase) if \ce{CO2} and \ce{H2O} are considered altogether. This corresponds respectively to a 7.7\si{\watt/\metre\squared} radiation blockage (\ce{CO2} alone) or 3.8\si{\watt/\metre\squared} radiation blockage (\ce{CO2}+\ce{H2O}). The examination of Fig \ref{fig:blackbody_transmitted} (bottom), hints at the mechanism that explains this increase of the absorption band: The line centers of the \ce{CO2} bands won't absorb any more radiation (since essentially they absorb 100\% of the incoming radiation, however as \ce{CO2} concentrations increase, so will the absorbing intensity of the wings, farther from the line center. Then this will lead to more radiation absorption on the edges of the spectral window where \ce{CO2} absorbs radiation, thereby effectively increasing the spectral range of the blockage. This is what is seen in the bottom part of figure \ref{fig:blackbody_transmitted}\footnote{We note that we have smoothed the simulation results for enhancing the visualization of this mechanism}.

One may then find it surprising on a first approach to focus on \ce{CO2} as the main greenhouse gas, since it is evident that \ce{H2O} is responsible for a larger percentage of absorption. The key difference lies in the fact that \ce{CO2} is a very stable molecule that may remain in the atmosphere for centuries, whereas water vapour can condense into liquid water and precipitate to the ground. On a first approach water vapour concentrations on the atmosphere will only depend on the temperature, which influences the dew point. Any excess of anthropogenic emissions of water vapour will simply condense back to liquid water. Not so much with \ce{CO2} emissions. This also brings in the question of feedback mechanisms: As temperature rises due to larger radiation blockage from \ce{CO2}, temperatures will rise, therefore increasing the percentages of water vapour, hence leading (on a first approach) to a positive feedback where increased water concentrations will block even more of the Earth's surface radiation. On the other hand, this will lead to an increase of clouds and an increase of Earth's albedo which will act as a negative feedback mechanism, as more of the Sun's radiation will be reflected back to Space. This is just a small example of the complexity for the different feedback mechanisms.

In conclusion, this is a simple, yet accurate analysis which may be run in any contemporary laptop computer in a matter of seconds. 
It shows the relevance of increased concentrations of \ce{CO2} in blocking radiative cooling from the Earth surface to Space, which has been neglected in Smirnov analysis. 

We note that part of the absorbed radiation will be re-emitted (equiprobably towards the Earth, or Space), where it may escape to Space/Earth surface or be absorbed again. One may refer to \cite{Wilson:2012} for the detailed treatment. This has not been carried out here, purely for the sake of simplicity.

\subsection{Possible modeling enhancements}

The analysis of Smirnov, as well as ours, has considered the simple case of radiative transfer in a perpendicular direction to Earth's surface, assuming a homogeneous column of air, and the isotropy of Blackbody radiation from Earth. This last assumption is confirmed by other authors 
\cite{Tang:2009,Warren:2019}, however, the former simplifications induce a few errors: For example if we assume a perpendicular direction for radiative transfer, we are underestimating the length for the column of air for the rays more tangential to the Earth's surface (as seen at sunset when only radiation from the red part of the sun's spectrum is transmitted). One may then carry out a more exact ray-tracing calculation using \cite{Vollmer:2006}:

\begin{equation}
\frac{h'(\theta)}{h}=\left[-\frac{R_E}{h}\cos(\theta)+\sqrt{\left(\frac{R_E}{h}\cos(\theta)\right)^2+2\frac{R_E}{h}+1}\right]
\end{equation}

where $h$ is the altitude of Earth's atmosphere column, $R_E$ is the radius of the Earth, and $\theta$ is the ray angle compared to Earth's surface normal vector.\\

Another improvement includes accounting for the difference in temperature and pressure in the air column. As the altitude increases, the density will decrease and the Lorentz broadening mechanisms will be less marked. Furthermore, Doppler broadening will also decrease as the result of lower temperatures. This will lead to spectral lines closer to diracs, with obvious impact on radiative transfer. However Wilson notes \cite{Wilson:2012} that simulations accounting for both these improvements yield similar results to the more simplistic approach, which means that likely these two simplifications cancel out to a given degree (ignoring tangential rays underpredicts radiation absorption, and ignoring lower line broadening at higher altitudes overpredicts absorption.

\section{Concluding Remarks}

We have shown that the influence of doubling the concentration of \ce{CO2} in the atmosphere on the surface temperature is not $\Delta T=0.02\si{\kelvin}$ as claimed by Smirnov, or even $\Delta T=0.4\si{\kelvin}$ if the correct fraction of the anthropogenic sources is considered for the doubling of \ce{CO2} concentrations in the atmosphere (100\% instead of 5\%). Only if one considers the influence on this doubling of \ce{CO2} concentrations on the energy flux $J_E$ that escapes from Earth, may one obtain a more correct value lying within 
$\Delta T=1.1-1.3 \si{\kelvin}$ \cite{Wilson:2012}.

Atmospheric heating from anthropogenic \ce{CO2} is a topic with very high societal impact, and as such should not be treated lightly. The last decade has brought a large wealth of new, large scale experimental and modeling works that have significantly reduced the uncertainties of several key mechanisms of global warming, namely feedback mechanisms \cite{Ciais:2014}. Some uncertainties remain, namely regarding the effects of aerosols, and the future trends of climate change, however, as illustrated in section \ref{sec:application}, modeling improvements and available computational power make it a trivial exercise to develop simple models that may evidence the key role of \ce{CO2} regarding the imbalance of Earth's energy budget, obviously excluding any feedback mechanisms, as a zero-order approach.

This is not to say that climate warming may no longer be considered complex topic, least one forget that the different feedback mechanisms to be accounted quickly complexify the necessary models and calculations, and that there are further issues with significant societal impact that need addressing, such as ocean acidification from increased concentrations from \ce{CO2}; the increase in re-circulation in the atmosphere (with the increase of extreme weather phenomena) resulting from the ocean temperature increase; or the impact on ocean currents.

\end{document}